# Structure and Magnetism in Mn Doped Zirconia: Density-functional Theory Studies

**Xingtao Jia[a], Wei Yang[b],*, Minghui Qin[c], Jianping Li[d]**

[a] *College of Chemistry & Chemical Engineering, China University of Petroleum, Dongying, 257061, China*
[b] *College of Information, Shanghai Ocean University, Shanghai, 200090, China*
[c] *National Laboratory of Solid State Microstructures and Department of Physics, Nanjing University, Nanjing 210093, China*
[d] *Department of Precision Instruments, Shandong College of Information Technology, Weifang, 261041, China*

**Abstract:** Using the first-principles density-functional theory plane-wave pseudopotential method, we investigate the structure and magnetism in 25% Mn substitutive and interstitial doped monoclinic, tetragonal and cubic $ZrO_2$ systematically. Our studies show that the introduction of Mn impurities into $ZrO_2$ not only stabilizes the high temperature phase, but also endows $ZrO_2$ with magnetism. Based on a simple crystal field model, we discuss the origination of magnetism in Mn doped $ZrO_2$. Finally, we discuss the effect of electron donor on the magnetism.

## 1. Introduction

Like many others tetravalent transition metal (TM) oxides, zirconia ($ZrO_2$) shows outstanding refractory property and finds application as structural ceramics, thermal barrier coatings, nuclear fuels, nuclear waste ceramics, solid electrolytes for oxide fuel cells and sensors, catalysts, alternate gate dielectrics and so on. Mainly, there are three crystal structures for $ZrO_2$. At ambient temperature, $ZrO_2$ shows monoclinic structure and transforms to tetragonal structure (> 1400 K) and then to cubic fluorite structure (> 2600 K) as temperature increases. For pure $ZrO_2$, the large stress induced by the volume change during the phase transition would destroy the material thoroughly. A traditional solution to the problem is the introduction of trivalent or divalent cations such as $Y^{3+}$, $Sc^{3+}$, $Sm^{3+}$, $Ca^{2+}$ [1-5], which would stabilize the high temperature phases to room temperature. Compared with the monoclinic $ZrO_2$ (*m*-$ZrO_2$) and tetragonal $ZrO_2$ (*t*-$ZrO_2$), the cubic $ZrO_2$ (*c*-$ZrO_2$) is a kind of good oxygen-transporting electrolyte in solid oxide fuel cells (SOFC) for the loose-packed lattices. It also finds use as oxygen sensors in steel making and automobile exhausts. Moreover, c-$ZrO_2$ show higher dielectric constant than monoclinic and tetragonal phase, and draw more attention as the key insulating gate material in complementary metal-oxide-semiconductor field effect transistors [6-8].

Recently, some studies show that *c*-$ZrO_2$ can be stabilized to room temperature by the introduction of Mn cations [9,10]; and theoretic studies demonstrate high temperature half-metallic ferromagnetism (HMF) in TM impurities doped *c*-$ZrO_2$ [11,12]. Compared with the traditional charge-based semiconductor technology, the semiconductor-spintronics paradigm is advantageous in using both the charge and spin of electrons at the same time, which has been considered to be a solution to the current silicon-based semiconductor's dilemma. The spintronics-based device is considered to be more powerful, endurable and not susceptible to radiation damage, and energy efficient than semiconductor-based one [13-15]. The ideal spintronics materials should own high temperature HMF, good compatibility with the mainstream semiconductor, and good synthesizability. Three types spintronics materials have been developed: magnetic metal and alloys, Heusler half-metals, and magnetic semiconductors. Among them, magnetic semiconductors are more promising for good compatibility with the current semiconductor technology, and ideal HMF character when doped with appropriate magnetic atoms. Benefiting from the development of modern molecular beam epitaxy (MBE) technology, the synthesis of magnetic semiconductors becomes more feasible. Besides HMF character, another advantage for the $ZrO_2$ based spintronics materials would be good compatibility with both group IV semiconductors and metals.

---

* Corresponding author.
E-mail address: yangwei_upc@yahoo.com.cn (W. Yang)

In the previous studies, we demonstrated robust HMF in Mn doped $c$-$ZrO_2$ [16]. Here, on the basis of the state-of-the-art *ab initio* electronic structure calculations, we investigate the structure and magnetism in Mn substitutive and interstitial doped $m$-, $t$- and $c$-$ZrO_2$ systematically, to find out the effect of Mn impurities on the structure evolution and magnetism. The studies show that the introduction of Mn impurities not only stabilizes the high temperature phase, but also endow $ZrO_2$ with magnetism.

## 2. Computational details

We employed plane-wave pseudopotential density functional theory (DFT) calculation, where the spin-dependent generalized gradient approximation (GGA) is used for the exchange and correlation effects [17]. In the total energy calculations, Ultrasoft pseudopotentials proposed by Vanderbilt were used to describe the ionic potentials of TM atoms [18], and the exchange-correlation functional parameterized by Perdew, Burke, and Ernzerhof (PBE) [19]. To simplify the calculation, the 25% Mn substitutive or interstitial doped $m$-, $t$-, and $c$-$ZrO_2$ was used in the calculations. We check the effect of the cutoff energy and Monkhorst-Pack grid on the ferromagnetic stabilization energy (energy difference between antiferromagnetic and ferromagnetic states, $\Delta E_{AF}=E_{AFM}-E_{FM}$) for the 25% Mn substituted $c$-$ZrO_2$ with different cutoff energies (450, 650 and 800eV) and Monkhorst-Pack grids (6x6x6 and 8x8x8), the results show that the ferromagnetic stabilization energy difference is less than 0.001 eV. So, the cutoff energy of 450 eV and Monkhorst-Pack grid of 6x6x6 is used in all calculation to save the calculation expense. All cells were optimized with relaxation of both lattice parameters and atomic positions. The total energy was converged to $1.0\times10^{-5}$ eV/atom while the Hellman-Feynman force was $3\times10^{-2}$ eV/A in the optimized cells.

## 3. Results and discussion

The studies about the TM impurities doped $ZrO_2$ are mainly focused on the improvement of thermodynamics, dielectricity, and mechanical properties, while the magnetism induced by the TM impurities are almost neglected. Since the report about the $d^0$ magnetism in $HfO_2$ by Venktesan and coworkers [20], although the subsequent studies demonstrated the magnetism come from magnetic contaminant [21], materials scientists have become interested in the wide band-gap (WBG) magnetic insulators. Here, we investigate the structure and magnetism in Mn doped $ZrO_2$ systematically, to find out the effect of Mn impurities on the structure evolution and magnetism in $ZrO_2$.

Because of the smaller ionic radius of Mn than Zr cation, the cell volume of the 25% Mn substituted $ZrO_2$ is contracted in all three phases while the lattice parameters are dependent. The lattice constants are contracted proportionally in the 25% Mn substituted $c$-$ZrO_2$, while that of the 25% Mn substituted $m$-$ZrO_2$ distort severely and almost lose all symmetry. Interestingly, the 25% Mn substituted $t$-$ZrO_2$ shows an analogous cubic lattice (a=b=5.088 Å, c=5.099 Å), indicates that the introduction of Mn substitution would stabilize the tetragonal phase into cubic phase. For the 25% Mn interstitial doped $ZrO_2$, both the cell volume and lattice constant are expanded proportionally in the cubic and tetragonal phase. In $m$-$ZrO_2$, the introduction of Mn interstitial almost restructures the monoclinic cell into an analogous tetragonal cell as shown in Fig. 1. In pure $m$-$ZrO_2$, Zr is coordinated with seven O atoms, while the coordination number is changed to six by the introduction of Mn interstitial. In Fig. 1, we can see a distorted octahedron structure formed by six O atoms, where the Mn interstitial centralizes in O octahedron. In conclusion, Mn interstitial can stabilize the $m$-$ZrO_2$ into tetragonal phase, while Mn substitution stabilizes the tetragonal phase into cubic phase.

The calculated magnetic ground state, total and Mn magnetic moment, ferromagnetic stabilization energy, and defect formation energy in 25% Mn substitutive and interstitial doped $m$-, $t$-, and $c$-$ZrO_2$ are given in table 1. Therein the formation energy of defects or impurities X is defined as [22]:

$$E_f[X] = E[X] - E[bulk] - \sum n_i \mu_i \qquad (1)$$

where $E[X]$ is the total energy from a supercell calculation with defect X in the cell, $E[bulk]$ is the total energy for the equivalent supercell containing only bulk $ZrO_2$, $n_i$ indicates the number of atoms of type $i$ that have been added to ($n_i > 0$) or removed ($n_i < 0$) from the supercell when the defect is created, and $\mu_i$ are the corresponding chemical potentials of these species. Here, We calculated $\mu_{Mn}$ for the antiferromagnetic β-Mn ground state and corrected by –0.07eV/Mn (the energy difference between the ground state of α-Mn and β-Mn according to Hobbs and coworkers) [23].

From table 1, we can see the total magnetic moments of the 25% Mn substitutive and interstitial doped $m$-, $t$-, and $c$-$ZrO_2$ show integral value of 3.0 $\mu_B$, while the Mn magnetic moments are different greatly, indicating the great difference in the microscopic environment around the Mn atoms. Therein, the Mn substituted $c$-$ZrO_2$ show robust HMF and Mn substituted $t$-$ZrO_2$ is less robust; the Mn interstitial doped $c$-$ZrO_2$ shows weak ground state antiferromagnetism (AFM), while Mn interstitial doped $t$-$ZrO_2$ posses weak HMF. The ferromagnetic stabilization energy of the Mn substituted $m$-$ZrO_2$ is close to zero, demonstrating a paraferromagnetism (para-FM) character; while the Mn interstitial doped $m$-$ZrO_2$ shows weak ground state HMF, which is originated from the p-d hybridization of the Mn atom in the anomalous O octahedron field as shown in Fig. 1. In all samples, the defects formation energies show large positive value, indicates that high concentration Mn defects in $ZrO_2$ can not form spontaneously. Comparatively, the Mn interstitial in $m$-$ZrO_2$ shows lower formation energy than other defects. In addition, it should be remembered that the positive formation energy does not mean impossibility to introduce Mn into $ZrO_2$.

The spin-dependent density of states (DOS) of the 25% Mn substitutive and interstitial doped $m$-, $t$-, and $c$-$ZrO_2$ are given in Fig. 2. Apparently, the Fermi energy ($E_F$) of the substitutive and interstitial doping structures shifts to the valence and conduction band side of bulk $ZrO_2$, respectively. The addition of Mn interstitial not only provides magnetic moment to $ZrO_2$ but also free carriers, which would shift the $E_F$ to higher energy. For the Mn substituted $m$-$ZrO_2$, the $E_F$ tangent to the majority DOS; which reaches to the peak around -0.1 eV beyond the $E_F$ while the minority DOS almost keeps a zero value at the same time. This is a typical half-metallic ferromagnetic semiconductor (HMS) character. Compared with the Mn substituted $t$- and $c$-$ZrO_2$, both the majority and minority DOS of the Mn substituted $m$-$ZrO_2$ show a lot of small irregular peaks. This is consistent with the lower symmetry in the Mn substituted $m$-$ZrO_2$. Compared with the Mn substituted $m$-$ZrO_2$, the DOS of the Mn interstitial doped $m$-$ZrO_2$ show a minority HMF character with less peaks, this is owing to the improvement of symmetry via the introduction of Mn interstitial as shown in Fig. 1. The $t$-$ZrO_2$ can be considered as a distorted $c$-$ZrO_2$ along (001) direction. So, the Mn substitutive and interstitial doped $t$- and $c$-$ZrO_2$ show similar DOS respectively. There exists an obvious difference in the majority DOS around the $E_F$ between the Mn substituted $c$- and $t$-$ZrO_2$, the former is mixed with the DOS of bulk $ZrO_2$ while the latter stands alone in the band gap of bulk $ZrO_2$. Apparently, both the Mn substituted $c$- and $t$-$ZrO_2$ show typical HMF character. There shows two impurity DOS peaks above the valence band separated by a pseudogap in the minority-spin channel, which should be associated to the $e_g$-$t_{2g}$ splitting expected from a simple crystal field model (Fig. 3). Interestingly, the Mn substituted $m$-, $t$-, and $c$-ZrO$_2$ not only show HMF character around the $E_F$, which also show HMF character in the higher energy, this is very useful character when used in ballistic hot-electrons injection scheme [24].

In Fig. 3, we give the schematic description of the band splitting and $p$-$d$ hybridization of Mn in O octahedron ($O_h$) and cubic field. In the $O_h$ field, the $d$ orbitals of the magnetic atoms would split into higher double-degenerated $e_g$ orbitals and lower triple-degenerated $t_{2g}$. The cubic filed can be divided into twofold tetrahedral ($T_d$) field. Therein, the $d$ orbitals would split into higher triple-degenerated $t_{2g}$ orbitals and lower double-degenerated $e_g$. For the symmetry reasons, in the $T_d$ field, only the Mn-$t_{2g}$ orbitals hybridize with neighbor O-$p$, while Mn-$e_g$ almost keep unhybridized. The former would further split into a couple of triple-degenerated bonding and antibonding $p$-$d$ hybrids orbitals (the bonding $p$-$d$ hybrids orbitals are mainly came from O-p, while antibonding one came from Mn-$t_{2g}$). In the $O_h$ field, the Mn-$e_g$ orbitals hybridize with O-$p$, while Mn-$t_{2g}$ almost keep unhybridized. The former would further split into triple-degenerated bonding $p$-$d$ hybrids and double-degenerated antibonding $p$-$d$ hybrids orbitals (the bonding $p$-$d$ hybrids orbitals are mainly came from O-p, while antibonding ones came from Mn-$e_g$). The valence electrons would enter into the lower energy orbitals firstly and then higher ones. As shown in Fig. 3(a), the Mn in O $O_h$ field would show majority HMF around the $E_F$, and the minority gap is formed by the bonding $p$-$d$ hybrids orbitals and unhybridized Mn-$t_{2g}$. Similarly, Mn in O cubic field also shows majority HMF, but the minority gap is formed by the Mn-$e_g$ and the bonding $p$-$d$ hybrids orbitals. One thing should be mentioned; the DOS of the Mn interstitial doped $m$-ZrO$_2$ shown in Fig. 2 does not consistent well with the scheme in Fig. 3(a), this is due to the free electrons introduced by the Mn interstitial, which would shift the $E_F$ to the higher energy.

There are many factors affecting the magnetism in magnetic semiconductors such as the type of TM impurities, doping profiles and concentration, type and crystal structure of semiconductor matrix. Although most TM impurities would endow semiconductor matrix with magnetism, most studies are focused on Cr, Mn, Fe, Co and Ni for more unpaired $d$ electrons. Semiconductors with zinc blende-like and fluorite-like structures have been demonstrated to be the most promising candidates as the magnetic impurities hosts [11, 12, 25, 26]. Here, we just discuss the effect of the doping profiles. As shown in our studies on the Mn doped ZrO$_2$, Mn substitution and interstitial in $c$-ZrO$_2$ show robust HMF and weak AFM respectively, and that in $m$-ZrO$_2$ show para-FM and weak HMF respectively. That is, the magnetism is strong structurally dependent. By using appropriate synthesis method (such as molecular beam epitaxy, MBE) and epitaxial substrate, the HMF character can be achieved. Otherwise, we can see the electron donors also show great effect on the magnetism. By adjusting the type and concentration of electron donors, a modulated electronic structure would be resulted (as shown in Fig. 2). For example, we can modulate the magnetism by shifting the $E_F$ in the Mn substituted $c$-ZrO$_2$ via co-doping with electron donors. However, some electron donors would deteriorate the ferromagnetic stabilization energy and the stability of the crystal structure, which should be avoided.

## 4. Conclusion

Using the first-principles DFT-GGA method, we investigate the structure and magnetism in Mn substitution and interstitial doped $m$-, $t$- and $c$-ZrO$_2$ systematically. Our studies show that Mn interstitial would stabilize the $m$-ZrO$_2$ into tetragonal phase, while Mn substitution stabilizes the $t$-ZrO$_2$ into cubic phase. The introduction of Mn substitution would endow $t$- and $c$-ZrO$_2$ with robust HMF; while Mn interstitial in $t$- and $c$-ZrO$_2$ shows weak HMF and weak AFM respectively. Mn substituted $m$-ZrO$_2$ shows para-FM while Mn interstitial doped $m$-ZrO$_2$ shows weak HMF. Using the simple crystal field theory, we elucidate the origination of HMF in Mn doped ZrO$_2$. Finally, we discussed the effect of electron donor on the magnetism.

**Acknowledge**

The work is partially supported by the Postgraduate Innovation Foundation of China University of Petroleum (No. B2008-8), and Xingtao Jia acknowledges the technological support of Wen He at the Shandon Institute of Light Industry.

**Table 1.** Calculated magnetic ground state, total and Mn magnetic moment ($m_{tot}$ and $m_{Mn}$), ferromagnetic stabilization energy ($\Delta E_{AF}$), and defect formation energy ($E_f$) in 25% Mn doped monoclinic, tetragonal, and cubic $ZrO_2$. Therein $Mn_{Zr}$ and $Mn_i$ respect Mn substitution and interstitial respectively.

|  | Defects | Ground state | $m_{tot}$ ($\mu_B$) | $m_{Mn}$ ($\mu_B$) | $E_{FA}$ (meV) | $E_f$ (eV) |
| --- | --- | --- | --- | --- | --- | --- |
| monoclinic | $Mn_{Zr}$ | Para-FM | 3.0 | 3.0 | 0.1 | 6.806 |
|  | $Mn_i$ | HMF | 3.0 | 3.7 | 9.1 | 3.976 |
| tetroganal | $Mn_{Zr}$ | HMF | 3.0 | 3.1 | 84.4 | 7.330 |
|  | $Mn_i$ | HMF | 3.0 | 3.3 | 8.1 | 4.688 |
| cubic | $Mn_{Zr}$ | HMF | 3.0 | 3.7 | 173.1 | 7.745 |
|  | $Mn_i$ | AFM | 3.0 | 3.7 | -3.4 | 4.638 |

**Figure captions**
Fig. 1 Scheme of optimized cell of Mn interstitial doped *m*-ZrO$_2$.
Fig. 2 Spin-dependence density of states (DOS) of 25% Mn doped ZrO$_2$. Therein the left column is Mn substitution and right Mn interstitial in ZrO$_2$, the panels from up to down are Mn impurities in *m*-, *t*-, and *c*-ZrO$_2$.
Fig. 3 Schematic description of the band splitting and *p-d* hybridization of Mn in O octahedron (a) and cubic (b) field. Therein, $d_1…d_5$ denote $d_x$, $d_y$, $d_z$, $d_{x^2-y^2}$ and $d_{z^2}$ respectively; and $p_1$, $p_2$, $p_2$ denote $p_x$, $p_y$, and $p_z$.

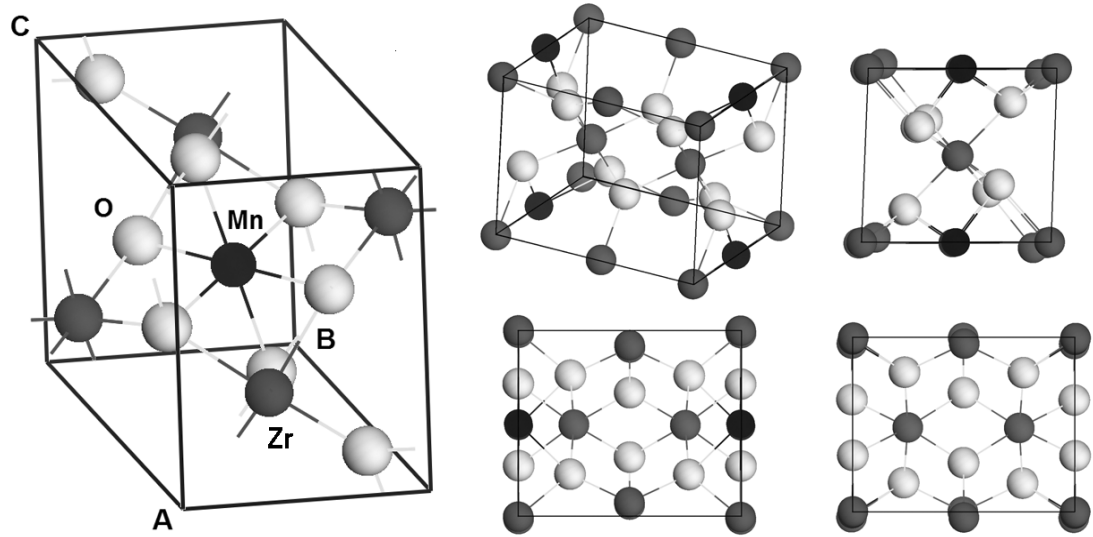

Fig. 1 X. Jia et al.

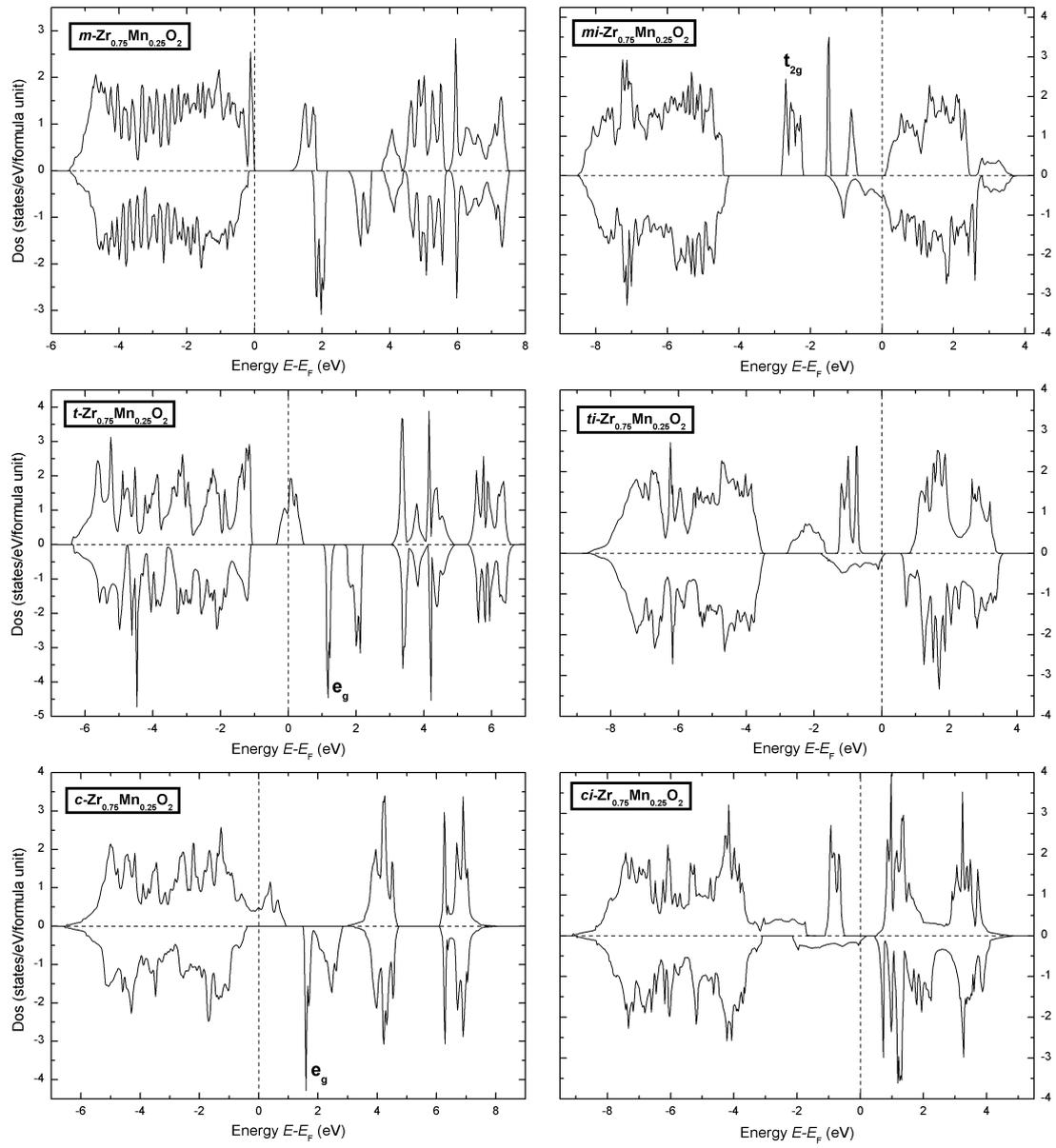

Fig. 2 X. Jia et al.

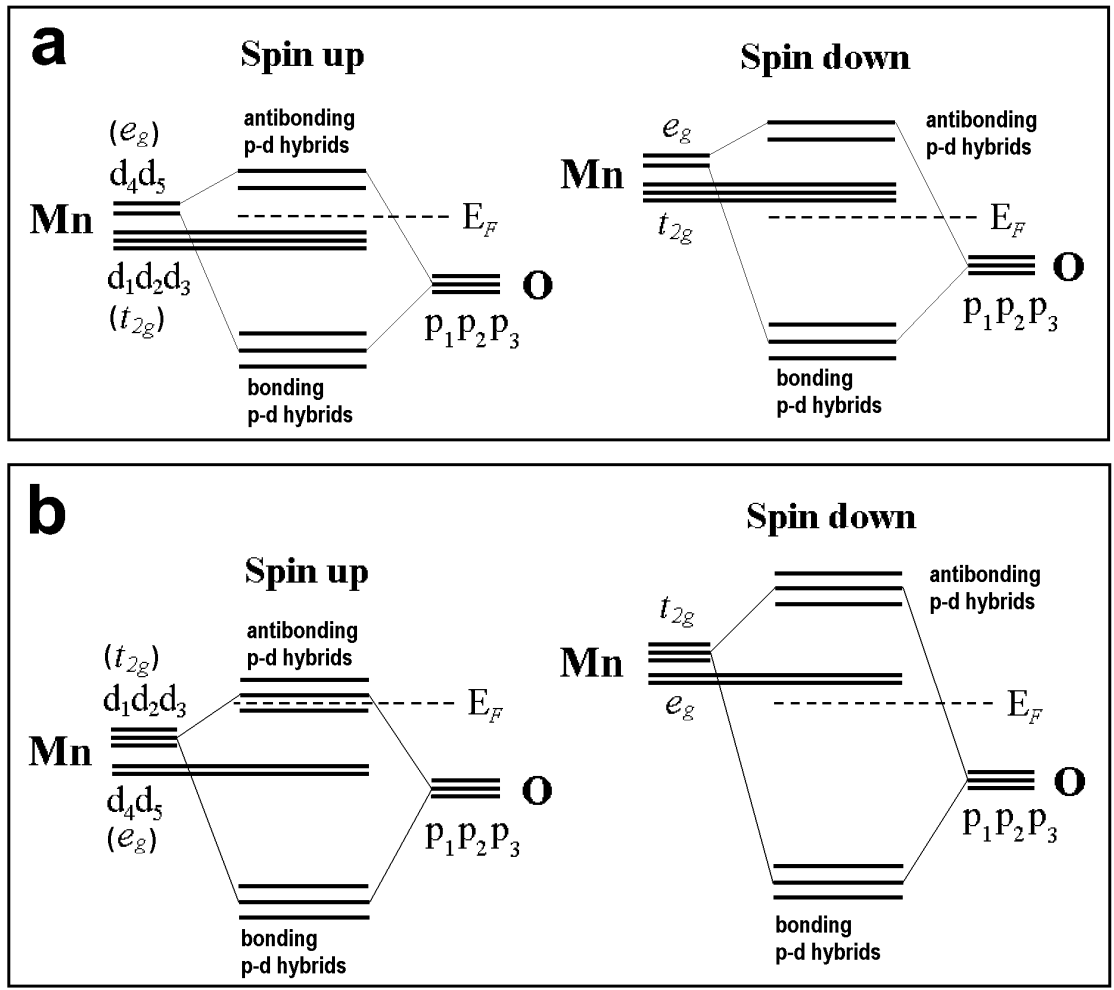

Fig. 3 X. Jia et al.